# The Development of Detectors of High-Energy Neutrinos and Cosmic Rays between 1980 and 2000

**A A Watson, School of Physics and Astronomy,**

**University of Leeds, Leeds LS16 5NU, UK**

a.a.watson@leeds.ac.uk

**Introduction:** Currently, information about astrophysical neutrinos above ~100 TeV is dominated by results from the IceCube detector at the South Pole, while data on cosmic rays above ~1 EeV come from the Pierre Auger Observatory located in Argentina and the Telescope Array located in the USA. These projects were initiated between 1980 and 2000 and attracted the attention of presidents of three countries. Claims in the 1980s for the detection of γ-rays above 1 PeV from Cygnus X-3, and from other X-ray binary systems, also had some impact on their evolution.

Unlike in particle physics where theory often guides experiment, predictions of fluxes of neutrinos and ultra-high energy cosmic rays were few in the early days and were generally unreliable as the sources of these particles remain a matter for considerable speculation. Indeed, identifying their sources remains a major goal. Thus, when seeking funding, the question 'Why do you want to make it so big?' is difficult to answer. In the 1980s and 1990s, many predictions of neutrino fluxes were little better than guesses. The situation for the highest-energy cosmic rays was more favourable as successively larger observatories had detected particles of increasing energy – so they were known to exist.

**The development of neutrino telescopes:** Moisej Markov (1960) [1] proposed that a neutrino telescope, consisting of detectors distributed over 1000 $m^2$, be deployed in a deep lake or in the sea, with Cherenkov radiation being used to detect the direction of charged particles produced by neutrinos.

In the late 1970s, Markov's idea was taken up by a group of American, European, Japanese and Soviet scientists led by Arthur Roberts who developed a proposal to deploy a detector at a depth of 5 km off the coast of Hawaii. It was envisaged that strings of photomultipliers would form the DUMAND (Deep Underwater Muon and Neutrino Detector) Array with signals transmitted via cable to the shore. After the Soviet invasion of Afghanistan in December 1979, it was made clear, prompted by President Ronald Reagan, that while the US scientists 'were free to choose their collaborators, funding would not be forthcoming if perchance they included Russians'(Roberts 1992 [2]).

In 1983, after several workshops, site surveys and technical tests, funding was released by the DOE and the NSF in the US, the University of Bern in Switzerland and the ICRR in Japan to deploy a prototype string of seven photomultipliers. In 1987 the string was deployed from a high-stability US naval vessel at depths between 2000 and 4000 m for a total of 38 hours, enabling the depth-dependence of the muon intensity to be measured accurately (Bobson et al. 1990 [3]). This proved to be one of the most important physics results obtained by the DUMAND team.

Three strings of photomultipliers are necessary to reconstruct a muon trajectory. In December 1993 the DUMAND group attempted such a deployment with a data link to shore from a junction box located on the ocean floor but problems with the pressure housings and a short circuit in the junction box meant that the effort was unsuccessful. These issues, together with the earlier deployment difficulties and the DOE's heightened sensitivity to failure following cancellation of the SSC, led to termination of the project. The DUMAND adventure, however, left a rich technical legacy, including the manufacture of large area photomultipliers by Hamamatsu and the use of optical fibres for data transmission.

**Lake Baikal Neutrino Telescope:** Even before the disconnection of Soviet scientists from DUMAND, the idea of instrumenting Lake Baikal as a neutrino detector was proposed by Alexander Chudakov (1979, [4]), with a dedicated laboratory funded at the Institute for Nuclear Research. Lake Baikal is 1700 m deep with exceptionally clear water and early in the year the ice thickness is sufficiently great to allow the use of heavy units for deployment of strings of photomultipliers more easily than from a ship.

Despite many difficulties, including the collapse of the Soviet Union in 1991, the team at Lake Baikal, enhanced with scientists from DESY (Zeuthen) led by Christian Spiering, achieved the first deployment of three strings in 1993. Thirty-six photomultipliers were used and, in 1997, the Lake Baikal team reported the first detections of upward-going, neutrino-induced events (Belolaptikov et al. 1997 [5]). Although the relatively shallow depth of Lake Baikal limits shielding from down-going muons, a strong program continues with 0.6 km$^3$ instrumented in 2025 and completion of a 1 km$^3$ detector foreseen in 2028.

**Development of IceCube:** In 1987 Francis Halzen learned of attempts at the Soviet Vostok base in Antarctica to detect neutrinos using the radio emission predicted from particle showers created by neutrinos in the ice. However, rough estimates suggested that only neutrinos of extremely high-energy would be detectable with the instrumentation then available. Instead, Halzen conceived the idea of using clear ice as the Cherenkov medium, with photomultipliers deployed in deep holes made using hot-water drilling. After a successful test of the concept in Greenland (Lowder et al. 1991 [6]) by a team led by Bob Morse, four strings, each carrying 20 photomultipliers, were deployed at the South Pole at depths between 800 and 1000 m as the AMANDA (Antarctic Muon and Neutrino Array) detector during the 1993/4 Antarctic summer. Almost immediately problems were identified within the ice volume surrounding the photomultipliers when it was found that light propagated between them more slowly than anticipated. An unexpectedly high rate of coincidences between air showers triggering the SPASE air-shower array, deployed in 1987 for other purposes, and muon signals in AMANDA led to the same conclusion and it was deduced that there were bubbles in the ice at the deployment depth. The Americans had been alerted to the possibility of air bubbles at these depths through a telex from Markov to John Learned, but the warning was overlooked (Bowen 2017 [7]). The high density of bubbles made reconstruction of muon directions impossible.

The SPASE array had been constructed to search for 100 TeV photons from X-ray binaries and SN1987A. This exploration stemmed from a claim made in 1983 for the observation of PeV photons from the X-ray binary system, Cygnus X-3, by a group at the University of Kiel (Samorski and Stamm 1983 [8]), a result apparently confirmed by the UK team working at Haverah Park (Lloyd-Evans et al. 1983 [9]). Other groups reported detections of PeV γ-rays from several similar systems and, after Michael Hillas (1986 [10]) pointed out that the best place to study such objects was the South Pole, a collaboration was formed between scientists from Bartol Research Institute, University of Delaware, and the University of Leeds. The detection of SN1987A and the subsequent predictions (Gaisser, Stanev and Halzen 1988 [11], Berezinsky and Ginzburg 1987 [12]) of detectable signals of high-energy photons from it enhanced the scope of the project.

The South Pole has several advantages as a location of an air-shower array. At an atmospheric depth of 690 g $cm^{-2}$, showers from relatively low-energy primary particles and photons can be detected and, being on the rotation axis of the earth, sources are visible at the same elevation for 24 hours each day. SN1987A is at an elevation of 21$^o$. However, the Pole presents a hostile environment with average winter temperatures of -60 $^o$C. Scintillators, made at MIT in the 1950s and loaned to the project by John Linsley, were mounted in light-tight boxes and viewed from below with a photomultiplier (figure 1, LH). In laying out the 16 detectors of the array, the access port to the photomultiplier and associated electronics was positioned so that the operator was shielded from the katabatic wind caused by cold air flowing down the slope of the glacier under gravity (figure 1, RH).

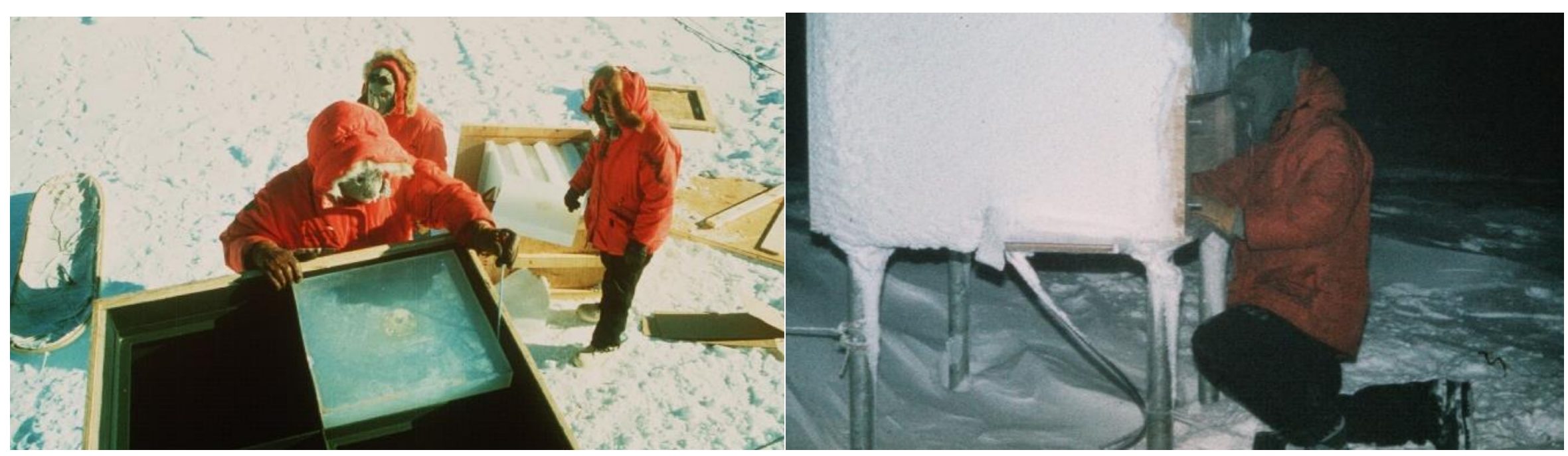

**Figure 1:** LH: A block of scintillator being loaded into one of the light-tight boxes. RH: The first winter-over scientist, Nigel Smith, working on one of the detectors during the winter of 1988. Smith is now Executive Director and CEO at TRIUMF (LH: A A Watson, personal collection: RH: N J T Smith, personal collection).

Observations started in December 1987 and the equipment worked flawlessly throughout the winter with over $10^6$ events recorded in the first two months of operation. These data were recovered from magnetic tapes only at the end of the winter as there was no high-speed internet at this time. No signals were seen from SN1987A with limits set well below predictions (Gaisser et al. 1989 [13]). The first SPASE array, and another positioned closer to AMANDA in 1994, were used to survey the positions of the AMANDA photomultipliers (Aherns et al. 2004 [14]). Additionally the mass of the cosmic rays around 1 PeV (Ahrens et al. 2004 [15]) was estimated using the 1996 deployment of AMANDA-B with 86 photomultipliers on four strings at depths between 1500 and 2000 m. Following the successful operation of AMANDA-B, the first IceCube modules were deployed between 2004 and 2005 at depths between 1450 and 2450 m, with an air-shower array located directly above. The IceCube Observatory was completed in the 2011-12 Antarctic season and has made ground-breaking detections of high-energy neutrinos up to energies of 10 PeV.

More detailed accounts of the development of the DUMAND and Lake Baikal detectors are given in papers by Roberts [2] and Spiering (2012 [16]). A fascinating account of the development of the IceCube detector in a less formal manner has been given by Bowen [7].

**The impact of Cygnus X-3:** The reports of high-energy $\gamma$-ray signals from Cygnus X-3 drew several eminent particle physicists to the cosmic-ray field. Undoubtedly the most prominent was James (Jim) Cronin who, with Val Fitch, had won the Nobel Prize in 1980 for the discovery of CP violation. Cronin outlined his reasons for his move from particle physics in an autobiographical article (Cronin 2014 [17]):

"*For 30 years, I performed experiments at high-energy accelerators. But the opportunity to work in a small group with a few students and senior colleagues was coming to an end. By the mid-1980s, many of the most important experiments proposed at accelerators were becoming larger and more expensive and often required hundreds of physicists………the evolving nature of the experiments did not appeal to me. Important as these future experiments were, I realized that they would take place with or without me. For any future experiments in which I would participate, I wanted to make a difference even if the experiment was not of the greatest importance. Thus I began to look for other areas of research.*"

Cronin carried out his final experiment in particle physics in 1985, while on sabbatical at CERN, leading a small group to measure the lifetime of the $\pi^0$ directly. The $10,000 budget made it possibly one of the cheapest experiments carried out there.

Intrigued by the reports of signals from Cygnus X-3 and frustrated by the low rate of detections because of the low-budget air shower arrays available, Cronin designed CASA (Borione et al. 1994 [18]), an instrument covering 1 $km^2$ which, if the initial reports [8, 9] were correct, would have detected several photons during each daily transit of the source across his array.

Jim Cronin was a modest man and wished to have his project critiqued by members of the cosmic ray community. Amongst other places, he visited Yakutsk in Siberia and in November 1986, *en route* to a CERN Council meeting, he asked to come to Leeds to spend a few days with my group. We interacted well and over the next few years a strong friendship and extensive scientific collaboration developed that continued until his death in 2016. That his father-in-law came from Edinburgh, my birthplace, and that we shared a love of malt whisky undoubtedly helped. The manner in which Cronin moved into the cosmic ray field, recognising that he had much to learn, contrasted rather sharply with others before him who had made a similar transition.

**Birth of the Pierre Auger Observatory**: By the mid-1980s, the rate of arrival of cosmic rays with energies above 10 EeV had been measured to be about one per $km^2$ per year. Particles with energies above 100 EeV were claimed to exist, with a flux of less than one per $km^2$ per century. In 1990 I gave a review talk on the status of the field (Watson 1991 [19]). My conclusions were somewhat downbeat:

*"It is disappointing and salutary to realise that these conclusions are not essentially different from those reached ten years ago. They are firmer now but it is depressing to think that the rate of collection of data is unlikely to allow very much stronger statements to be made within the next five years. The problem is lack of exposure: while it has been clear for many years that 1000 $km^2$ of instrumental area is needed, progress towards achieving this has been slow.*

*There are tantalising indications to be followed up and some theoretical predictions to be tested. The experimental problems are challenging and subtle but certainly soluble. All that is needed is dedication, money and patience; otherwise, a reviewer writing ten years from now will not be able to draw many different conclusions."*

I sent the written version of my talk to Jim and when we next met at the International Cosmic Ray Conference in Dublin in August 1991, his first words to me, before any greeting, were

*"You're not nearly ambitious enough: we should build 5000 km$^2$!"*

Particle physicists think big, just as Jim had done when planning the CASA array, and this greeting was the start of a journey that led to the creation of the Pierre Auger Observatory, the largest cosmic ray detector ever constructed. During Jim's four-month sabbatical in Leeds in late 1991, we made the first plans for the Observatory.

**Building the Collaboration:** We recognised that we had to build a large collaboration to achieve our ambitions, and it was an opportune time to attract particle physicists, as well as cosmic-ray physicists, to the project. The SSC had been terminated in October 1993 and several physicists decided to leave the LEP project at CERN, including Giorgio Matthiae, well-known for his role in the development of the Roman pots used at the ISR.

At the insistence of Don Perkins, we also approached CERN. Don knew that the CERN charter states that its activities should include the study of cosmic rays, a provision probably put there at the insistence of Pierre Auger and Eduardo Amaldi. Our approaches to Chris Llewellyn-Smith, the Director-General at the time, were disappointing: CERN was under financial pressure because of the LHC and no immediate help was forthcoming. However, thanks to Chris and Lorenzo Foa, CERN later agreed to 'recognise' the project and to act as a holding bank for 'common funds'. This was extremely valuable: contributions to these funds came from each country and there was reluctance for the monies to be held in Argentina, whilst the Argentinians would not countenance Fermilab as the holding place.

Having signed up many scientists from Europe, the USA and Latin America, we turned our attention to the Far East, making a visit to six countries in three weeks in 1994. Our first stopping point was Tokyo as we were eager to engage the very talented people working on the AGASA project (Chiba et al. 1992 [20]), the 100 km$^2$ array that was beginning to take data at the Akeno site near Tokyo. However, we found the Japanese group, then led by Masahiro Teshima, to be unenthusiastic about joining a relatively large collaboration, believing that the desired aperture could be achieved using the method of detecting the fluorescence radiation emitted from nitrogen molecules in the atmosphere, excited by the charged particles in the air showers. Essentially the atmosphere acts as a calorimeter and detecting the fluorescence light holds the promise of measuring the primary energy directly. The method had been developed as the Fly's Eye (Baltrusaitis et al. 1985 [21]) by a team from Utah and all expertise in this area lay with them. The Japanese view proved to be a handicap when seeking US funding and, in hindsight, was undoubtedly a poor decision. The fluorescence technique is challenging for many reasons, not least because the material of the calorimeter, the atmosphere, is a variable which requires intensive monitoring.

A memorable part of our trip was our visit to Vietnam with which the USA had no diplomatic links: Jim travelled with a special visa. However, we were warmly welcomed and met Nguyen Van Hieu, the Vice-President of the Communist Party, a theoretical physicist who had been in Dubna when Jim had visited shortly after the discovery of CP violation in 1964. Van Hieu, a Lenin Prize winner, had been a research student of Markov. I watched, fascinated, as these two 'elder statesmen' talked of those days in an elegant reception chamber where the following evening, on TV, we watched Indira Gandhi being greeted.

Our trip was largely unsuccessful from the viewpoint of increasing the strength of the collaboration although physicists from Vietnam, under the direction of Pierre Darriulat who had been the spokesperson for UA2, later became involved in the project. The Adelaide group, led by Roger Clay and Bruce Dawson, had a long history in cosmic rays and were keen to host the project at the Woomera rocket range that we had visited during an early Workshop in 1993 but by the time of this trip possible difficulties over Aboriginal land rights had surfaced. Also, there was little prospect of strong financial support.

The decision to name the Observatory after Pierre Auger was taken in recognition of his major role in unravelling several of the properties of air-showers in the late 1930s in France, and later in Chicago, where he was based before moving to Canada to work on topics related to the Manhattan Project.

**The Design Study:** In the first six months of 1995, a Design Study was held at Fermilab with the strong support of the Director, John Peoples. He had recently read a book in which the question of the source of cosmic rays was raised and was convinced that we were planning important work. Cronin was able to extract \$100k from UNESCO to support scientists from developing countries to take part in the Design Study, for necessary travel, and for a site search to take place. Jim was always able to get through doors that I could not even have knocked on, and this was a particularly striking example as in 1995 the US was not a member of UNESCO.

The design study resulted in a proposal to build a hybrid detector, with an array of water-Cherenkov detectors overlooked by a set of telescopes to detect the fluorescence emission from the air showers. A schematic of a portion of the instrument is shown in figure 2.

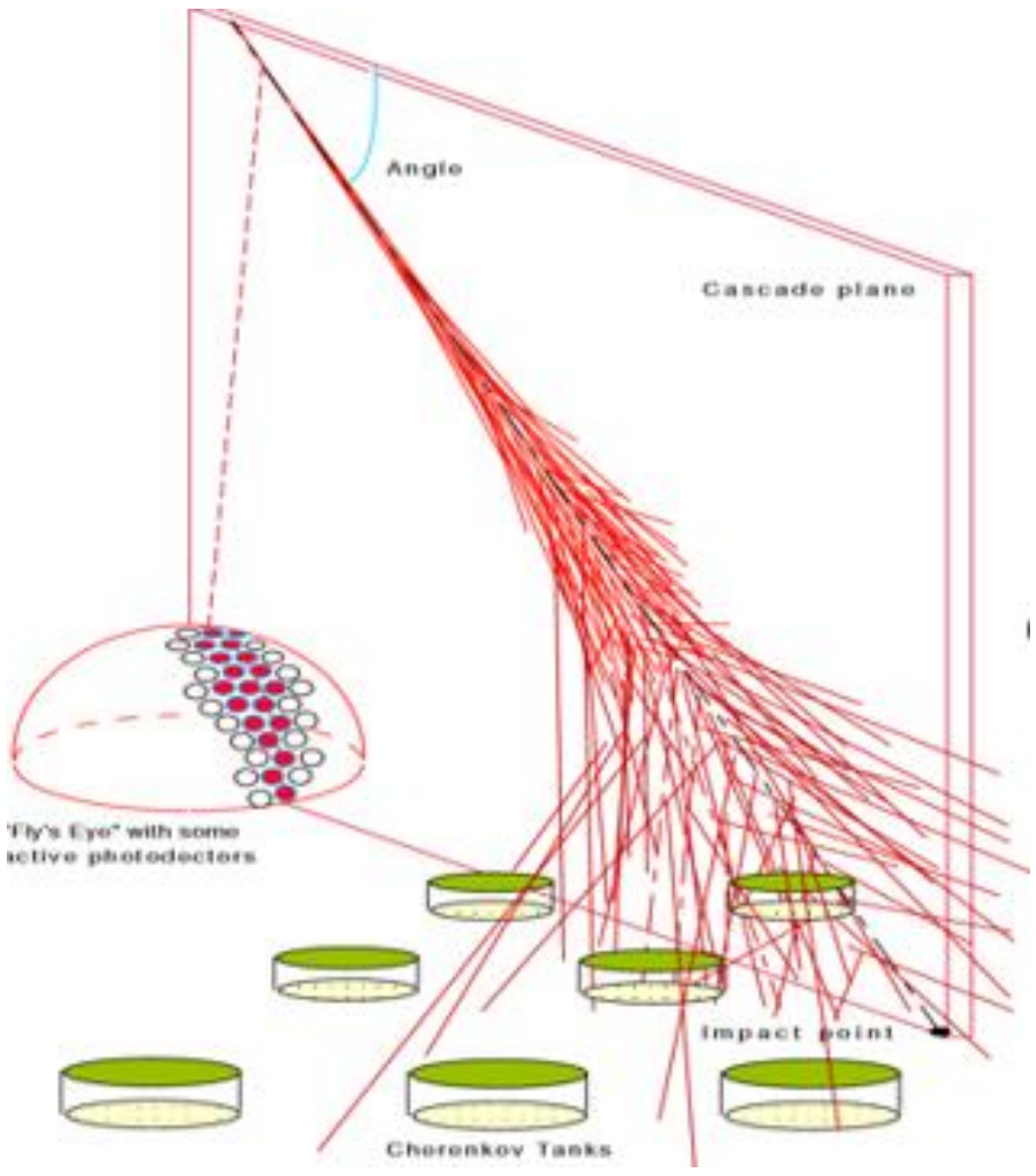

Figure 2: Schematic of the layout with some water-Cherenkov detectors overlooked by a fluorescence detector. The Observatory contains 1600 water-Cherenkov detectors covering an area of 3000 km$^2$, overlooked by four fluorescence telescopes. (Diagram by Enrique Zas)

The plan was to construct observatories in the Northern and Southern Hemispheres. Here the depth of the water-Cherenkov detectors gives a significant advantage over scintillators as detection of events at very inclined angles to the zenith is practical. It had been demonstrated at Haverah Park that water could be kept clean, without filtration, for many years: a tank that was opened at the end of the project which had contained untreated water for 25 years is shown in figure 3. The white material, DARVIC, designed by ICI for use in sandwich boxes, had an anti-bacterial content that inhibited growth of anything that would have reduced the clarity of the water. Cronin liked to describe this photograph as 'an existence proof'. A rough figure of what we thought we might raise was \$100M, sufficient to construct two arrays each covering 3000 km$^2$.

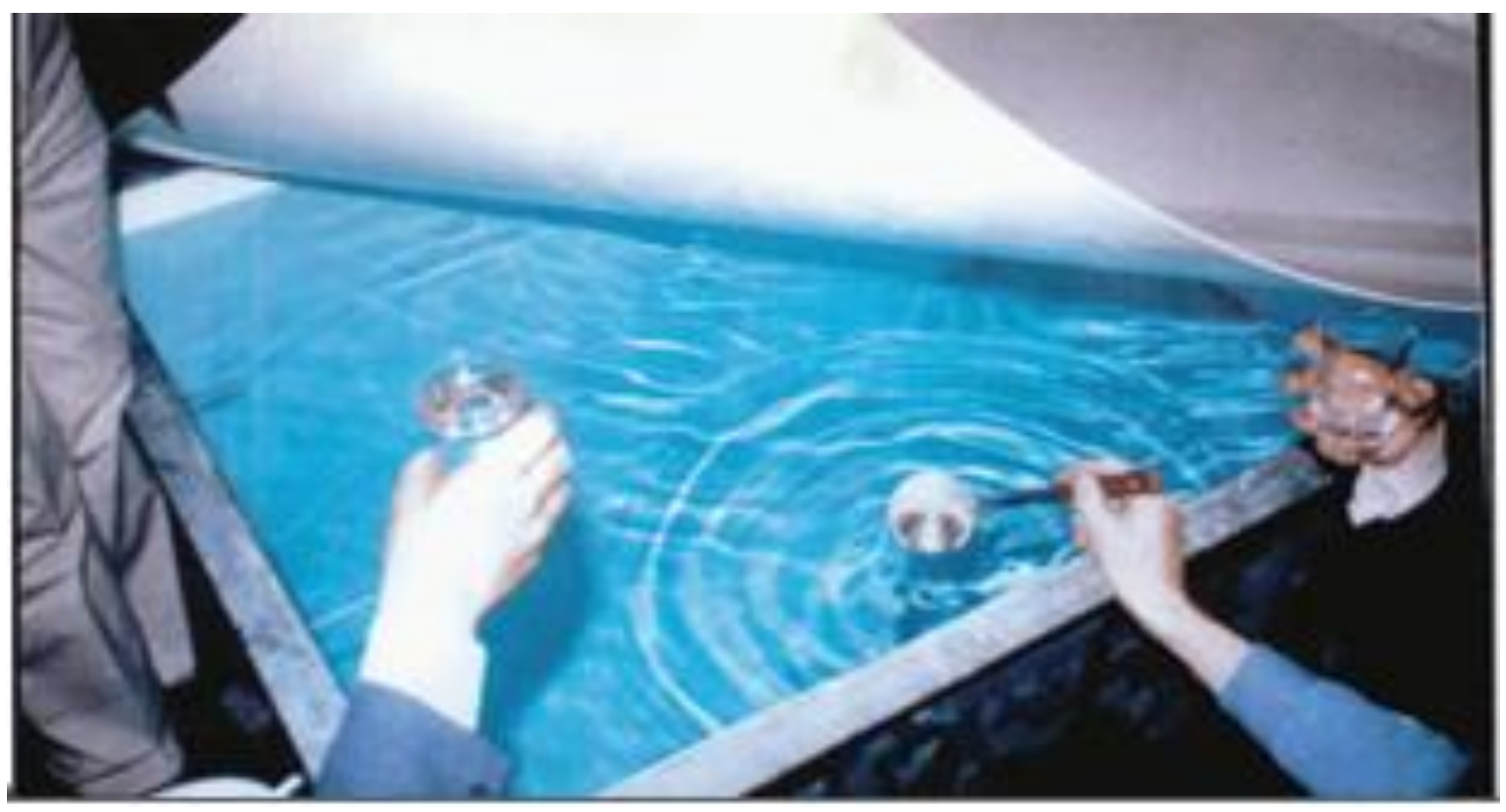

**Figure 3:** A water-Cherenkov tank, which had contained untreated water for over 25 years, that was opened at the closure of the Haverah Park project in July 1987 (A A Watson (personal collection)).

The success of the Design Workshop, in which several members of Fermilab contributed strongly, raised the possibility that the Lab might play a significant role. Accordingly Peoples commissioned a review evaluation carried out by a panel that included John Bahcall, Francis Halzen and Tom Gaisser. The supportive report from this group was not favourably received by the Lab's Program Advisory Committee who were unhappy with Peoples' proposal to support the project. In the end, he chose to ignore the advice of the Advisory Committee, a move which led to significant hostility from many in the Fermilab community. Peoples' decision led to key members of the Collaboration coming from Fermilab, including Paul Mantsch, a first-class project manager, and Peter Mazur, who played a major role in the design and construction of the water-Cherenkov tanks.

**Site Searches:** To give all-sky coverage with two arrays, sites at latitudes between 30$^{o}$ and 40$^{o}$, north and south were needed, at an altitude between 400 and 1500 m, with clear skies, relatively flat and with suitable local infrastructure. Financial support for what would be the host country was also seen as essential. Searches were undertaken by Ken Gibbs and Antoine Letessier-Selvon who wore out their passports exploring possibilities in the Northern Hemisphere in the western United States, Mexico, Spain, Russia and Kazakhstan, and in the Southern Hemisphere where the only possible locations were in Australia, South Africa and Argentina. The possibility in South Africa was on land owned by only two landowners, and we had a strongly encouraging letter of invitation from President Nelson Mandela. However, there was no prospect of financial support and political stability in South Africa was far from certain as Mandela had been released from prison only five years earlier. By contrast, President Carlos Menem offered 11 M Ars Pesos to attract us to Argentina, where there were several suitable sites. At this time, the Argentinian peso and the US dollar had parity. One of the appeals to Menem appears to have been the opportunity to be photographed alongside Cronin. Menem was attempting to get the constitution changed so that he could run for a third term and seemed to think that these photo-ops would help his case. His attempt failed, but in a reciprocal move

Cronin arranged to have the Auger Observatory on the agenda when President Clinton met the new Argentinian President, de la Rúa, in Washington in 2000. In 2002, the peso was devalued to 4:1 against the US dollar. But the Argentinians kept their word, and the project still got 11 M Ars Pesos, leaving a shortfall which other agencies made up.

**A Review Panel:** There was no obvious route to have the plans of the Collaboration, set out in the 250-page Design Report (Pierre Auger Collaboration 1985 [22]), assessed so Cronin took the unusual step of inviting an international panel to assess our plans and to grill the proponents. The *ad hoc* six-member Panel, chaired by Ian Axford, a distinguished cosmic-ray theorist and Max Planck Director at Lindau, included Ron Ekers, Jack Steinberger and Masatoshi Koshiba. The Panel returned a very supportive report which was used effectively to gain funding in several countries. In the US, however, the response from the National Science Foundation (NSF) was '*Of course, it was a strong report: you chose the Panel'*.

**Assessment of the Project in the USA:** It was always clear that the project was unlikely to proceed without substantial funding to Jim Cronin with the other partner countries looking to the US funding agencies to show serious commitment. A major difficulty was that since 1974 NSF had been supporting a group, based largely at the University of Utah, for studies of the highest-energy cosmic rays using fluorescence detectors and consequently were rather unreceptive to our novel approach. The Utah group had never shown much interest in supplementing their fluorescence work with an array of surface detectors and indeed an effort to develop this concept during a week-long meeting in 1979 ended in failure. They, along with the Tokyo team, were convinced that an array of fluorescence detectors would provide a lower-cost alternative to achieving the same aperture. Furthermore, the US agencies were keen to maintain control and paid little regard to the existence of a strong international collaboration. The US proposal for $30M was considered by a sub-committee (SAGENAP), the Scientific and Advisory Group for Experiments in Non-Accelerator Particle Physics. At its first meeting in March 1997, SAGENAP decided to defer the recommendation providing a set of questions to the Collaboration.

Presentations at the following meeting of SAGENAP eight months later set out to answer these questions in detail. However, towards the end of the meeting a young scientist from the Utah group, Dave Kieda, presented a novel idea for detecting the highest-energy cosmic rays. Named 'El Cheapo', the proposal was to use solar panels to detect Cherenkov light from showers with energies above 1 EeV. Although the idea had been published (Kieda 1995 [23]), (I was a receiving editor for the journal at the time and had expedited its publication), there was no experimental demonstration that the method would be effective. I reacted so strongly to such a blatant effort to undermine our detailed and well-reviewed plans that Cronin was advised that it might be best if I did not attend subsequent SAGENAP meetings. The decision of the Committee was to reject the proposal (it is said by a single vote). They did, however, endorse the science goals, asked us to descope and to optimise the use of fluorescence detectors.

After much soul searching, the US groups returned with a proposal for $15M with the hope that the other collaborators would be able to make up the shortfall. DOE were generally more supportive than NSF who did not like the descoped plans as there was no proposal to collaborate with the groups in Utah and Japan. However, the revised application led to a $7.5M award in July 1998 for four years with construction to begin in Argentina but with the prospect of a Northern site postponed to the indefinite future. Cronin was devastated by this outcome and needed significant persuasion not to abandon the project. However, we went ahead and funding from other countries came in relatively quickly leading to a ground-breaking ceremony in Argentina in March 1999. This was attended by the daughter of Pierre Auger who is shown with Jim Cronin in figure 4.

Although motivated for the wrong reasons, the decision to go to Argentina proved to be a wise one. It would have been too difficult to have constructed two sites simultaneously with the team available.

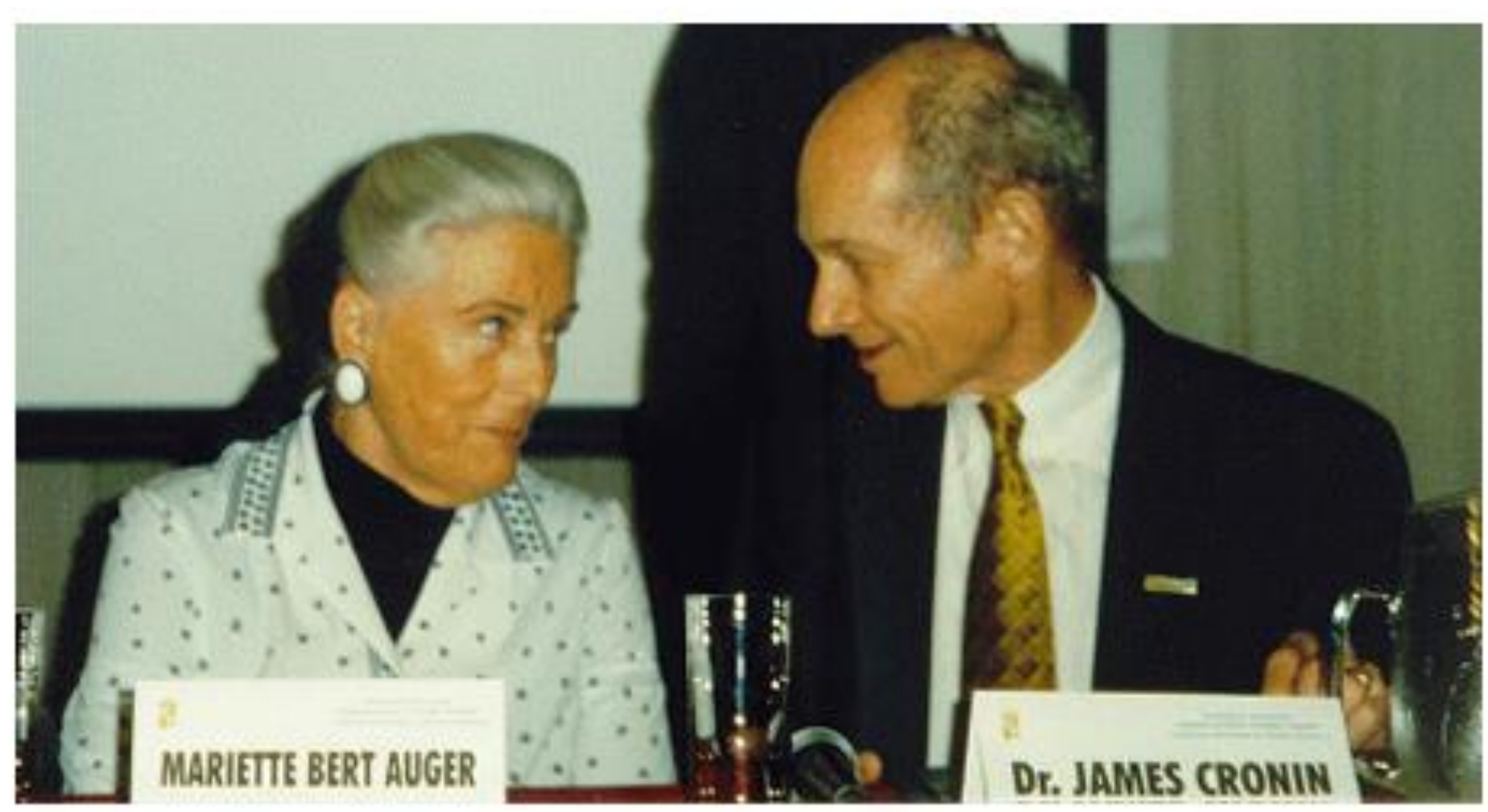

**Figure 4:** Jim Cronin with Marlette Berl Auger (the name plate is incorrect) during the ground-breaking ceremony in Malargüe, Argentina in March 1999 (Pierre Auger Collaboration collection).

Some years after the ‘El Cheapo’ incident, during the International Cosmic Ray Conference in Mexico in 2007, where I reported the first results from our completed detector, Kieda apologised for his intervention, making it clear that he had been asked to do this by senior colleagues in Utah. He was, at the time, approaching tenure review. The unpleasant incident led him to move from studying high-energy cosmic rays to γ-ray astronomy.

Cronin persisted with attempts to draw the Utah group into the project. In late 1998 he came out of retirement to take up a five-year faculty position at the University of Utah. But attempts to engage the senior members of the Utah group failed and he resigned after a year. In the wake of his visit and early departure, several talented younger people left Utah for faculty positions elsewhere and became key members of the Auger Collaboration.

**Main Physics Questions:** The only clear prediction about the highest-energy cosmic rays had been made in 1966 by Greisen (1966 [24]) and by Zatsepin and Kuzmin (1966 [25]) following the discovery of the cosmic microwave background – the GZK effect. They pointed out that interactions between protons and the photons of the 2.7 K radiation field would excite the Δ-resonance and lead to an abrupt steepening of the cosmic-ray spectrum above about 50 EeV. For heavy nuclei, photoproduction would lead to a similar suppression. Arrays operating until the late 1980s were too small to test this prediction but the 100 km$^2$ AGASA detector, which began operation in 1990, soon claimed the observation of particles with energies well-beyond the GZK prediction. Numerous explanations were proposed including the breakdown of Lorentz invariance (Coleman and Glashow 1999 [26]) and that the particles might be evidence of cosmological relics (Berezinsky 2000 [27]). In addition, an event of 300 EeV was reported from the HiRes detector, the successor to the Fly's Eye instrument (Bird et al. 1995 [28]). These events gave added impetus to the need to build a detector with a large aperture.

A further enhancement of the physics potential of the Auger Observatory was the realisation that it could be used to detect very high-energy neutrinos (Capelle et al. 1998 [29]). At large angles to the zenith, the electromagnetic component of a shower produced by a nuclear primary would be largely absorbed leaving predominantly muons. The showers would be characterised by a flat shower-front and a narrow spread in the arrival time of the shower particles. By contrast, a shower created by a neutrino could be initiated deep in the atmosphere and would look very much like a shower arriving in a near-vertical direction. The distinction is illustrated in figure 5 where the signals from a very inclined event are shown and contrasted with those from a near-vertical event.

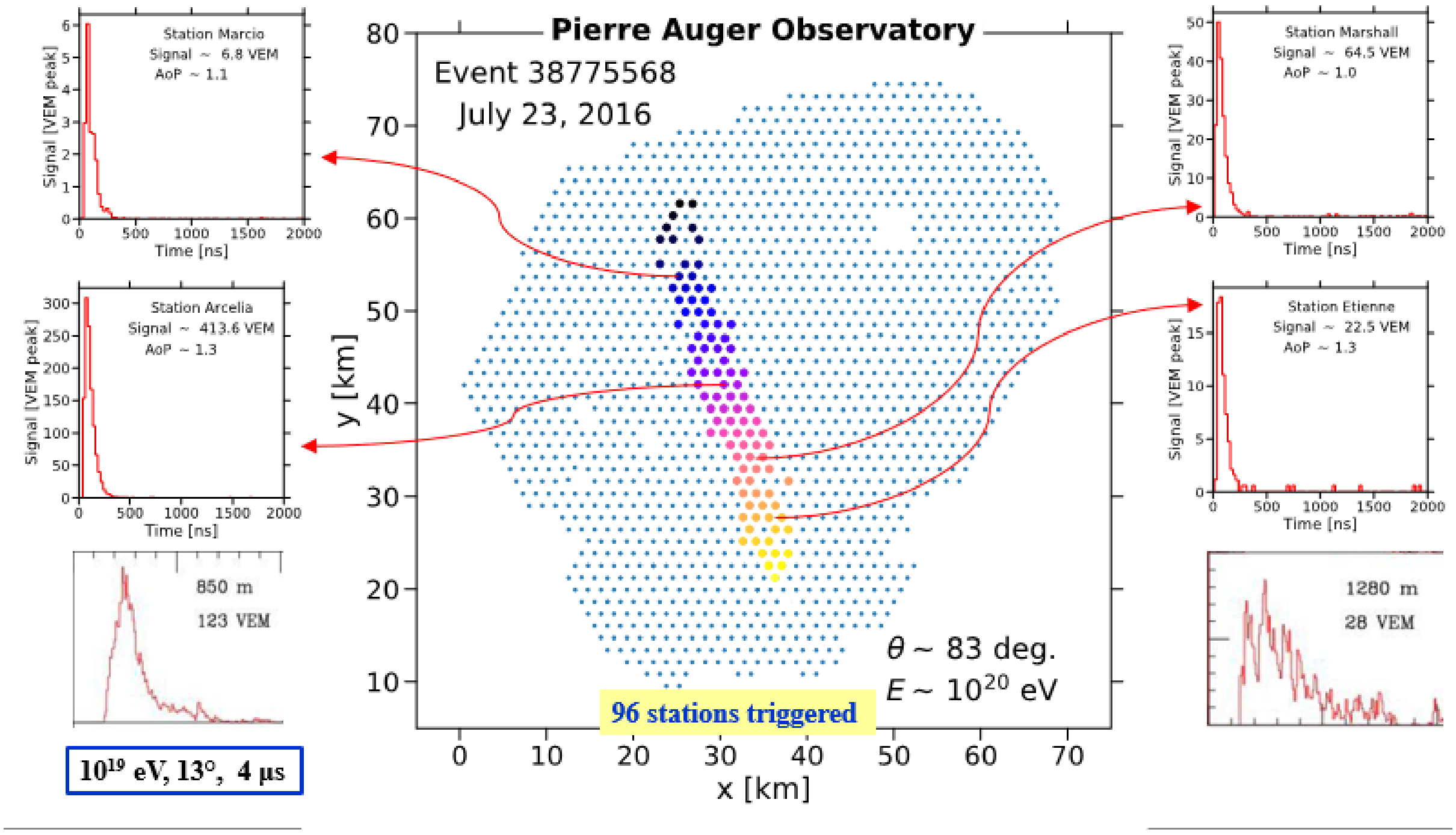

**Figure 5:** An event at large zenith angle in which signals were recorded in 96 water-Cherenkov detectors. The narrow time spread of the signals in four of the detectors is shown. By contrast, the two lower plots are for a near-vertical event where a large time spread is evident. Signals like these, but seen at large zenith angles, would be expected in a shower initiated by a neutrino. (Pierre Auger Collaboration collection).

**The Telescope Array:** The realisation that the hybrid approach to studying showers was superior to a large array of fluorescence detectors eventually led to the formation of a joint US-Japanese collaboration (Telescope Array) in 2003. Under the leadership of Pierre Sokolsky and Masahiro Fukushima, 507 scintillators arranged on a 1.2 km grid, overlooked by three fluorescence detectors, were laid out over 700 km$^2$ in Utah. Operation began in 2007.

**Concluding Remarks:** While operating arrays of water-Cherenkov or scintillator detectors is relatively straightforward, with on-times in excess of 95% readily achieved, the same cannot be said of fluorescence detectors. Operation is restricted to clear, moonless nights and, in addition, there is the need to maintain the condition of the mirrors and to monitor the atmosphere to continuously to determine the aerosol content and the cloud cover. Weather variability means that the collection of high-quality data over long periods is challenging.

Also, the efficiency of conversion of electromagnetic energy into fluorescence photons remains relatively uncertain. It is clear, for example, that the low levels of anisotropy of cosmic rays above 4 EeV discovered by the Auger Collaboration (The Pierre Auger Collaboration 2017 [30]) would have been impossible to detect with fluorescence detectors alone.

The period from 1980 to 2000 led to the construction of devices that have revolutionised our knowledge of astrophysical neutrinos and of the highest-energy cosmic rays. In the following decades, operation of IceCube, the Lake Baikal detector and the Auger Observatory have been supplemented by efforts in the Mediterranean with the ANTARES and KM3NeT neutrino detectors and Telescope Array to study high-energy cosmic rays respectively. Other neutrino detectors are under development in the China Sea (TRIDENT) and off the western coast of Canada (P-One), while IceCube-Gen2 will provide a volume of 8 $km^3$. The Auger Observatory continues to be used to search for neutrinos above 1 EeV.

The enigmatic story of Cygnus X-3, which had a small role in the early days of neutrino searches at the South Pole, and a major role in the development of the Pierre Auger Observatory, has taken a further twist with the recent report by the LHAASO group (Zhen Cao et al. 2025 [31]) that Cygnus X-3 can be quiescent for several years. In active periods, photons of PeV energies were detected in coincidence with observations at lower energies from the Fermi LAT detector.

**Acknowledgements:** I am grateful to the organisers for inviting me to speak at the Symposium and to the Royal Society (London) for financial support.